\documentclass[11pt,letterpaper]{article}
\usepackage{graphicx}
\usepackage{slashed}
\usepackage{latexsym}
\usepackage{amssymb}
\usepackage{amsbsy}
\usepackage{epsfig}

\title{Decays of the $B_{c}$ Meson in a\\Relativistic Quark--Meson Model}
\author{Matthew A. Nobes \\ 
Department of Physics, Simon Fraser University\\ Burnaby, BC, Canada V5A 1S6\\ 
and \\ R. M. Woloshyn\\
TRIUMF, 4004 Wesbrook Mall, Vancouver, BC, Canada V6T 2A3 }

\date{}

\begin{document}

%\maketitle

\begin{flushright}
TRI-PP-00-20
\end{flushright}

\begin{center}

{\huge Decays of the $B_{c}$ Meson in a\\Relativistic Quark--Meson Model\\}

\vspace{0.5cm}

Matthew A. Nobes \\ 
Department of Physics, Simon Fraser University\\ Burnaby, BC, Canada V5A 1S6\\ 
and \\ R. M. Woloshyn\\
TRIUMF, 4004 Wesbrook Mall, Vancouver, BC, Canada V6T 2A3

\end{center}

\begin{abstract}
The semileptonic decay form factors of the double heavy $B_{c}$ meson provide a unique
opportunity to study the strong interactions between two heavy quarks.  
A fully relativistic model, with effective non-local quark-meson interactions, 
is used to compute semileptonic decay form factors, 
for both the $B_{c}$ and a wide range of other heavy-light mesons.
Using these form factors predictions for decay rates and branching ratios are 
obtained.
The results are compared to other theoretical approaches and, where available, 
to experimental results.  
In addition the radiative decay of $B^*_{c}$ is discussed.
\end{abstract}

\newpage

\section{Introduction} \label{sec1}

A primary goal in the study of semileptonic decays of heavy mesons
is to extract the values of the CKM matrix elements.
The great virtue of semileptonic decays is that the 
effects of the strong interaction can be 
separated from the effects of the weak interaction into a set of Lorentz invariant
form factors~\cite{gilmansingleton}. Thus the theoretical problem associated with analysing 
semileptonic decays is essentially that of calculating the form factors.

The focus of this work is the decay of the 
$B_{c}$ meson (for a review of the properties of this system see~\cite{gershteinetal}).
This system is unique among mesons made up of heavy (charm or bottom) quarks, 
it is the only one which is stable with respect to strong and electromagnetic
interactions. Therefore, the $B_{c}$ system is the only heavy meson
for which form factors (albiet transition form factors rather than elastic)
can be measured. These form factors then provide a unique probe of the 
dynamics of heavy quark systems.

There are many approaches to the calculation of decay form factors, 
for example, lattice QCD~\cite{latticerev}, QCD sum rules~\cite{bcsr1}, and phenomological 
modelling~\cite{WBS}. In this work a particular model with an effective
quark-meson coupling is adopted. There are many models of this type~\cite{ENJL1,
sutherlandlewisetal,blaschkeetal,deandrea,QCMbook}.
The one used here has its genesis in the QCD version of the Nambu-Jona-Lasinio
model~\cite{NJL}
extended to heavy quarks~\cite{ENJL1}
and is most closely
related to the model used recently by Ivanov and Santorelli in a their study of
pseudoscalar meson decays~\cite{ivanovsant}.

The advantage of this
approach is that it is fully relativistic and very versatile. Quarks and
mesons for all masses are treated within the same framework. For light quarks
the model has the features of spontaneous chiral symmetry breaking and in the
single heavy quark limit the form factor constraints of heavy quark 
symmetry are obtained.

Our work differs from Ivanov and Santorelli in the choice of the quark-meson
vertex function and in the way that parameters are fixed. A number of heavy
mesons decays not calculated in Ref.\cite{ivanovsant} are treated here. The main new
results are the extension of the model to include doubly heavy mesons, the calculation
of $B_{c}$ semileptonic decays and the electromagnetic vector to 
pseudoscalar transitions.

This paper is organized as follows: the next section introduces the 
model, discusses the general method of calculation, and fits the models free parameters.
Sect.~\ref{sec3} presents the calculation of the form factors and decay rates for the 
semileptonic decays of a wide varity of heavy-light pseudoscalar mesons.  These
calculations are compared with both measured results, and other theoretical
approaches. Sect.~\ref{sec4} presents
the same set of calculations for the eight primary semileptonic decays of the 
$B_{c}$ meson.  The predictions are compared with other theoretical work, in order
to highlight the differences that exist between various approaches.
Sect.~\ref{sec5} briefly discusses the electromagnetic decays $V \to P + \gamma$
for a number of vector mesons, including the $B^{*}_{c}$.  Sect.~\ref{sec6} gives
conclusions and directions for future work.

\section{Quark--Meson Coupling} \label{sec2}

The particular quark--meson coupling used in this work 
is based on an effective Lagrangian which models the
interaction between mesons and quarks with a non-local interaction 
vertex~\cite{ENJL1,ivanovsant}.
The interaction Lagrangian has the form

\begin{eqnarray}
L_{\mathrm{int}}(x) & = & g^{}_{M}M(x) \int\!\!dx_{1} dx_{2} \, \delta\! \left( x- \left( 
\frac{1}{2}\frac{m_{1}x_{1}+m_{1}x_{2}}{m_{1}+m_{2}} \right) \right) \times \nonumber \\
& & f[(x_{1}-x_{2})^{2}] \bar{q}_{1}(x_{1}) \Gamma_{M} q_{2}(x_{2}), \label{eq:lint}
\end{eqnarray}
where $\Gamma_{M}$ is the Dirac matrix appropriate to the meson field M, 
$f[(x_{1}-x_{2})^{2}]$ is a non-local vertex function, which simulates the 
finite size
of the meson, and $q_{1}$ and $q_{2}$ are the quark fields. A condition imposed on 
the  vertex function is that it should render all loop diagrams UV finite.
The coupling constant $g^{}_{M}$ is 
determined by the compositeness condition, which is the requirement that the 
renormalization constant of the meson fields be zero, i.{}e.

\begin{equation} \label{eq:cc}
Z_{M}=1-\frac{g_{M}^{2}}{2} \frac{d \Pi_{M}(p^{2})}{dp^{2}} \Bigg\vert_{p^{2}=M_{M}^{2}}
=0 .
\end{equation}
Here $\Pi_{M}(p^{2})$ is the self energy of the meson field, given by
\begin{eqnarray} 
\Pi_{M}(p^{2}) & = & 2N_{c} \int\!\! \frac{d^{4}k}{(2\pi)^{4}i} f^{2}(Q^{2}) \nonumber \\
& & \mathrm{tr} \left\lbrace \Gamma_{M} \frac{1}{m_{1}-(\slashed{k}+\slashed{p})}
\Gamma_{M} \frac{1}{m_{2}-\slashed{k}} \right\rbrace , \label{eq:self1}
\end{eqnarray}
where $m_{1}$ and $m_{2}$ are the masses of the quarks in the loop and $Q$ is a relative momentum
chosen to be $Q=k+\alpha p$ with $\alpha=\frac{m_{2}}{m_{1}+m_{2}}$.

The constituent quark masses in (\ref{eq:self1}) are free parameters. As well,
the vertex function will contain a free parameter which reflects the size of 
the meson.  These parameters will be different for the 
different mesons.

The use of free constituent quark propagators in expressions like (\ref{eq:self1})
can lead to a problem which reflects the lack of quark confinement in the model.
If the meson mass $M_{M}$ is greater than the sum of its constituent quark 
masses loop integrals
will develop imaginary parts.  This indicates a non-zero amplitude for the creation
of a free quark-antiquark pair.  There have been some various attempts to
obviate this problem within quark-meson effective 
theories~\cite{sutherlandlewisetal,blaschkeetal,deandrea,QCMbook}.  
Here we adopt the approach of Ref.\cite{ivanovsant} and use free propagators.
The constituent quark masses are then fit to allow
for the inclusion of as many mesons as possible.  

In order to carry out calculations a choice must be made for the vertex function 
$f(q^{2})$.
The function that was used in this analysis was the di\-pole
$$
f(Q^{2})=\frac{\Lambda^{4}}{[\Lambda^{2}-Q^{2}]^{2}}.
$$
This choice was made for two
reasons; first the form of the dipole vertex function is the same as a 
propagator, allowing standard Feynman parameter techniques to be used in evaluating loop
integrals.  Second, the vector decay constant $f_{V}$ would  
diverge if only  a monopole vertex function was used.  
Since one of the primary criteria for the 
vertex functions is that they should render all diagrams UV finite, a function   
with UV fall-off as least as fast as a dipole is needed.
The parameter $\Lambda$ characterizes the finite size of the meson, and will
be different for different mesons.  To account for this the various values of 
$\Lambda$ will be distinguished by subscripts which reflect either the meson type
or the quark content, \emph{e.{}g.}~$\Lambda_{B_{c}}$ and $\Lambda_{bc}$ will
be used interchangeably.  Further, in expressions involving the vertex form factor,
the same comvention will be used.
Note that the calculations of Ref.\cite{ivanovsant}
used a Gaussian vertex function so that the parameters used there can not be
compared directly with ours.  

The parameters of the model were fit to the leptonic decay constants, $f_{P}$ and 
$f_{V}$.
These quantities are defined by
\begin{eqnarray}
\langle 0 \vert -i\gamma^{\mu}\gamma^{5} \vert P \rangle & = & i f_{P} p^{\mu}, \\
\langle 0 \vert i \gamma^{\mu} \vert V, \epsilon \rangle & = & M_{V}^{2} f_{V} 
\epsilon^{\mu}
\end{eqnarray}
where $M_{V}$ is the vector meson mass.
The pseudoscalar decay constant is given by the one-loop expression
\begin{eqnarray}
f_{P} p^{\mu} & = & N_{c} \int\!\! \frac{d^{4}k}{(2\pi)^{4}i} g^{}_{P} f(Q^{2})
\nonumber \\
& & \frac{\mathrm{tr} \left\lbrace -\gamma^{\mu} \gamma^{5} [m_{1}+\slashed{k}+
\slashed{p}] \gamma^{5} [m_{2}+\slashed{k}] \right\rbrace }{ \left[ m_{1}^{2}-(k+p)^{2}
\right] \left[ m_{2}^{2}-k^{2} \right] } . \label{eq:pcons}
\end{eqnarray}
Here $m_{1}$ and $m_{2}$ refer to the masses of the quarks in the loop, 
this convention will be used throughout this paper.
Using the dipole vertex function, combining the denominators using Feynman parameters, and
performing the integration over k yields
\begin{equation} \label{eq:fp}
f_{P}=g^{}_{P} \frac{3 \Lambda_{P}^{4}}{4\pi^2} \int\!\! D\vec{x}
\frac{x_{1} [m_{2}(1-\sigma)+m_{1}\sigma]}{\Delta^2},
\end{equation}
with
\begin{eqnarray}
\sigma & = & \alpha x_{1} +x_{2} , \nonumber \\
\eta & = & \alpha^{2} x_{1} +x_{2} ,\nonumber \\
\Delta & = & \Lambda_{P}^{2} x_{1} + m_{1}^{2} x_{2} + m_{2}^{2} x_{3} +(\sigma^{2}-\eta)
M_{P}^{2} ,\nonumber \\
\ell & = & k+\sigma p , \nonumber \\
\int\!\! D\vec{x} & = & \int_{0}^{1} \left( \prod_{i=1}^{3} dx_{i} \right) \delta\!\!
\left( \sum_{i=1}^{3} x_{i} -1 \right) , \nonumber
\end{eqnarray}
where $M_{P}$ is the mass of the pseudoscalar meson.
Likewise the expression for the vector decay constant is
\begin{equation} \label{eq:fv}
f_{V}=g^{}_{V} \frac{3\Lambda_{V}^{4}}{4\pi^{2}M_{V}^{2}} \int\!\! D\vec{x}\,x_{1}
\frac{m_{1}m_{2}+\Delta+\sigma(1-\sigma)M_{V}^{2}}{\Delta^{2}},
\end{equation}
where the same defintions have been used, with the obvious change of $M_{P}$ to 
$M_{V}$ and $\Lambda_{P}$ to $\Lambda_{V}$ in the expression for $\Delta$.

To compute the coupling constants $g_{P}$ and $g_{V}$, the self energies and their
derivatives must be computed.  Then the compositeness condition (\ref{eq:cc}) can be
used to find the couplings.  The self energy for a pseudoscalar meson is given by
\begin{equation} \label{eq:selfp}
\Pi_{P}(p^{2})=\frac{3\Lambda_{P}^{8}}{2\pi^{2}} \int\!\! D\vec{x} \, x_{1}^{3}
\frac{m_{1}m_{2}+p^{2}\sigma(1-\sigma)+\frac{2}{3} \bar{\Delta}}
{\bar{\Delta}^{4}},
\end{equation}
where $\bar{\Delta}=\Lambda_{P}^{2} x_{1} + m_{1}^{2} x_{2} + m_{2}^{2} x_{3} 
+(\sigma^{2}-\eta)p^{2}$ and all the other quantities are the same as the ones
defined above.
The self energy for a pseudoscalar meson is given by the tensor $\Pi^{\mu\nu}_{V}$ which
can be expressed as
\begin{equation}
\Pi_{V}^{\mu \nu}(p)=\Pi_{V}(p^{2}) g_{\mu \nu} + \bar{\Pi}_{V}(p^{2})
\frac{p_{\mu} p_{\nu}}{p^{2}}. 
\end{equation}
Unfortunately $\Pi_{V} \ne \bar{\Pi}_{V}$, so this does not have the proper structure for
a vector propagator.  This problem was solved (following~\cite{heavyQCM1}) by simply dropping
the $\bar{\Pi}_{V}$ term,
which would cancel out of any calculation of a physical process at one-loop order (since 
$\epsilon \cdot p=0$).
The relevant part of the vector meson self energy is given by
\begin{equation} \label{eq:selfv}
\Pi_{V}(p^{2})=\frac{3\Lambda_{V}^{8}}{2\pi^{2}} \int\!\! D\vec{x}\, x_{1}^{3}
\frac{m_{1}m_{2}+\frac{1}{3}\bar{\Delta}+\sigma(1-\sigma)p^{2}}{\bar{\Delta}^{4}} ,
\end{equation}
where all the quantities appearing have been defined previously.

The free parameters of the model are fit to the six values of $f_{P}$ and the measured
values $f_{J/\psi}=0.1309$ and $f_{\Upsilon}=0.075012$~\cite{PDB}.  These data,
which are displayed in Table~\ref{fittedps}, fix eight free parameters.
In order to reduce the number of free parameters to match the available data the value
of the strange quark mass was fixed at 450 MeV and the vertex parameter for a vertex containing
only u and d quarks $\Lambda_{\pi}$ was taken (following~\cite{ENJL1}) to be 1 GeV.  In 
addition the following further simplifying assumptions were made
\begin{displaymath}
\Lambda_{us}=\Lambda_{ds}=\Lambda_{ss}=\Lambda_{K},
\end{displaymath}
\begin{displaymath}
\Lambda_{uc}=\Lambda_{dc}=\Lambda_{sc}=\Lambda_{D},
\end{displaymath}
\begin{displaymath}
\Lambda_{ub}=\Lambda_{db}=\Lambda_{sb}=\Lambda_{B}.
\end{displaymath}

This leaves the following parameters to be fit, $m_{q}$, $m_{c}$, $m_{b}$, $\Lambda_{K}$,
$\Lambda_{D}$, $\Lambda_{B}$, $\Lambda_{cc}$, $\Lambda_{bb}$, and $\Lambda_{bc}$.
The parameter $\Lambda_{bc}$ could only be fit to a value for $f_{B_{c}}$ which is not
in the values listed in Table~\ref{fittedps}, hence it is retained as a free parameter, leaving 
eight to be fit.  The fit to the remaining eight parameters is given by 
(all values in MeV)
\begin{eqnarray}
m_{u,d} & = & 245 ,\nonumber \\
m_{c} & = & 1800, \nonumber \\
m_{b} & = & 5100, \nonumber \\
\Lambda_{K} & = & 1225, \nonumber \\
\Lambda_{D} & = & 1350, \nonumber \\
\Lambda_{B} & = & 1500 , \nonumber \\
\Lambda_{cc} & = & 1420, \nonumber \\
\Lambda_{bb} & = & 2900. \nonumber 
\end{eqnarray}
The values for the self energies, coupling constants, and leptonic decay constants arising
from these parameters are displayed in Tables~\ref{pcalc} and~\ref{vcalc}.

In order to fix $\Lambda_{bc}$ a value of $f_{B_{c}}$ must be given.  There
is no experimental value for this quantity and theoretical estimates
tend to fall in the range 
$400\, \mathrm{MeV}  \lessapprox f_{B_{c}} \lessapprox 500\, \mathrm{MeV}$ 
(see, for example,~\cite{gershteinetal,eichtenquigg,colangelodefazio, 
anisimovISGW,joneswoloshyn}).
A further complication is that the mass of $M_{B_{c}}$ is also not yet 
known very well.  The current measurement~\cite{CDF} is
$M^{CDF}_{B_{c}}=6.4\pm 0.39(stat)\pm 0.13(sys)
\,\,
\frac{\mathrm{GeV}}{c^{2}}$
, which comes from
the few confirmed $B_{c}$ events at the Tevatron.
Theoretical results tend to lie within this range, so
following the potential model prediction of~\cite{gershteinetal} the mass of the 
$B_{c}$ was chosen to be 6.25 GeV.

\begin{figure}[!ht]
\begin{center}
\input{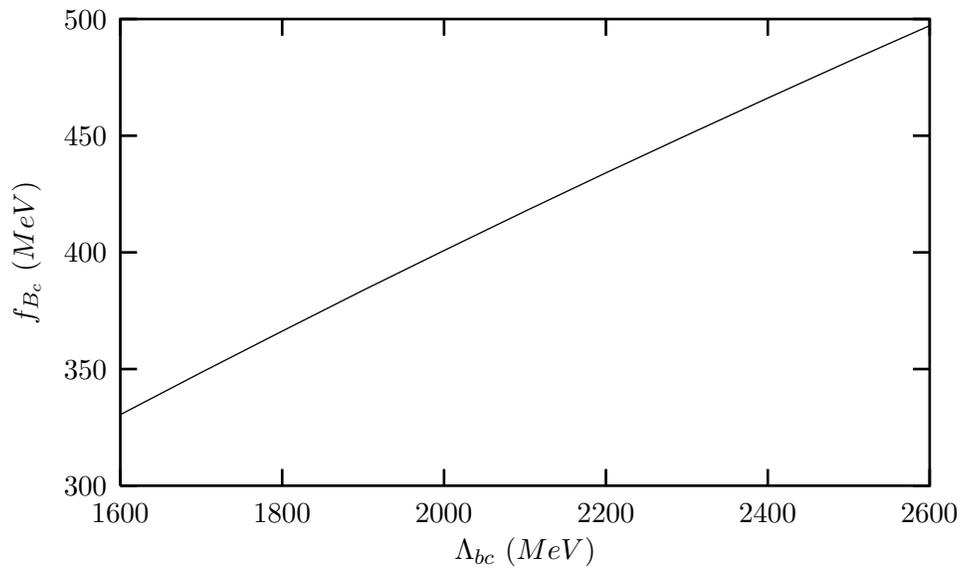}
\end{center}
\caption{$f_{B_{c}}$ vs. $\Lambda_{bc}$} 
\label{fbc}
\end{figure}

One general argument guides the selection of $\Lambda_{bc}$, it
should lie between $\Lambda_{B}$ and $\Lambda_{bb}$.  With this in mind, and using the
value for $M_{B_{c}}$ above, a number of values of $\Lambda_{bc}$ were tried, spanning the 
possible range.  Fig.~\ref{fbc} shows the value of $f_{B_{c}}$ as a function 
of $\Lambda_{bc}$. 
The value selected selected for use in this work was $\Lambda_{bc}=2.3$ GeV, which 
gives $f_{B_{c}}=450$ MeV, a value in the middle of the range of the theoretical
predictions.

\section{Semileptonic Decays of $K$, $D$, and $B$ Mesons} \label{sec3}

The model used in this work is phenomenological but having fixed its parameters,
the results for semileptonic decays are predictions. Before proceeding to
decays of $B_c$ it is important to test the model against experimental results
where they are available. Therefore 
several semileptonic decays of $K$, $D$, and $B$ meson
are calculated. The formalism for these calculations, presented in this
section, extends directly also to the calculation of $B_c$ decay.

Some of the decays considered here have already been treated by 
Ivanov and Santorelli~\cite{ivanovsant}. 
However, that work does not demonstrate the full applicablity of the approach.
Apart from decays to light vector mesons, the model is capable of treating virtually any
semileptonic decay (with the restriction that a value for the meson mass must be supplied
as input).

The amplitude $A$ for a semileptonic decay is given by,
\begin{equation} \label{eq:ampform}
A=\frac{G_{F}}{\sqrt{2}} V_{QQ'} L_{\mu} H^{\mu}.
\end{equation}
Here $G_{F}$ is the Fermi constant, $V_{QQ'}$ is the relevant CKM matrix element,
$L_{\mu}$ is the lepton current
\begin{displaymath}
L_{\mu}=\bar{u}_{\nu_{\ell}} \gamma_{\mu} (1-\gamma^{5}) v_{\ell} ,
\end{displaymath}
and $H^{\mu}$ is the hadron current
\begin{equation} \label{eq:hadcurr}
H^{\mu}=\langle k,\epsilon \vert (V^{\mu}-A^{\mu}) \vert P \rangle ,
\end{equation}
where $P$ is the momentum of the parent meson, $k$ is the momentum of the daughter meson,
and $\epsilon$ is the polarization, if the daughter meson is a vector.  The two currents 
in (\ref{eq:hadcurr}) are the vector $V^{\mu}$ and axial $A^{\mu}$.
If the final state is a pseudoscalar the hadron current can be decomposed as follows,
\begin{eqnarray}
\langle k \vert A^{\mu} \vert P \rangle & = & 0 ,\nonumber \\
\langle k \vert V^{\mu} \vert P \rangle & = & f_{+}(q^{2})(P+k)^{\mu}+
f_{-}(q^{2})(P-k)^{\mu}, \nonumber 
\end{eqnarray}
where $f_{+}(q^{2})$ and $f_{-}(q^{2})$ are Lorentz invariant form factors.
Likewise, if the final state is a vector meson,
\begin{eqnarray}
\langle k,\,\epsilon \vert A^{\mu} \vert P \rangle & = & f(q^{2}) \epsilon^{*\,\mu}
+a_{+}(q^{2}) (\epsilon^{*}\cdot P)(P+k)^{\mu} + \nonumber \\
& & a_{-}(q^{2}) (\epsilon^{*}\cdot P)
(P-k)^{\mu} , \nonumber \\
\langle k,\,\epsilon \vert V^{\mu} \vert P \rangle & = & ig(q^{2}) 
\epsilon^{\mu \nu \rho \sigma} \epsilon^{*}_{\nu} (P+k)_{\rho} (P-k)_{\sigma} ,
\nonumber
\end{eqnarray}
where the form factors are $g$, $f$, $a_{+}$, and $a_{-}$.
In each of these expressions $q=(P-k)$ is the momentum transfer.

For a decay to a pseudoscalar meson (with mass denoted by $M_{P'}$) the differential 
decay rate can be reduced to~\cite{gilmansingleton}
\begin{equation} \label{eq:dgdyps}
\frac{d\Gamma}{dq}=\frac{G_{F}^{2} \vert V_{QQ'} \vert^{2}
M_{P}^{2} K^{3}}{24 \pi^{3}} \vert f_{+}(q^{2}) \vert^{2} .
\end{equation}
where,
\begin{equation} \label{eq:K}
K=\frac{M_{P}}{2} \sqrt{\left[ 1-\frac{M_{P'}^{2}}{M_{P}^{2}}-y \right]^{2}
-4\frac{M_{P'}^{2}}{M_{P}^{2}}y}.
\end{equation}
The lepton spectrum is given by,
\begin{eqnarray} 
\frac{d\Gamma}{dx} & = & \frac{G_{F}^{2} \vert V_{QQ'} \vert^{2} M_{P}^{5}}{16 \pi^{3}}
(1-2x) \nonumber \\
& & \int_{0}^{y_{max}(x)} \left( [y_{max}(x)-y] \vert f_{+}(q^{2}) \vert^{2} 
\right) dy , \label{eq:pselec}
\end{eqnarray}
where $y_{max}(x)=\frac{4x(x_{max}-x)}{1-2x}$ with 
$x_{max}=\frac{M_{P}^{2}-M_{P'}^{2}}{2M_{P}^{2}}$.
If the final state is a vector meson (with mass $M_{V}$)the corresponding 
differential decay rate is,
\begin{equation} \label{eq:dgdyvec}
\frac{d\Gamma}{dy}=\frac{G_{F}^{2} \vert V_{QQ'} \vert^{2} KM_{P}^{2}y}{96\pi^{3}}
\left( \vert \bar{H}_{+}\vert^{2} + \vert \bar{H}_{-}\vert^{2} + 
\vert \bar{H}_{0}\vert^{2} \right) ,
\end{equation}
where
\begin{eqnarray}
\bar{H}_{\pm} & = & f(q^{2}) \mp 2M_{P}Kg(q^{2}) ,\nonumber \\
\bar{H}_{0} & = & \left[ \frac{M_{P}}{2M_{V}\sqrt{y}} \right]
\left[ \left( 1-\frac{M_{V}^{2}}{M_{P}^{2}}-y \right) f(q^{2}) +4K^{2}a_{+}(q^{2}) \right] , 
\nonumber
\end{eqnarray}
and the final mass $M_{V}$ should be subsititued for $M_{P'}$ in (\ref{eq:K}).
The expression for the lepton spectrum is given by
\begin{eqnarray} 
\frac{d\Gamma}{dx} & = & \frac{G_{F}^{2} \vert V_{QQ'} \vert^{2} M_{P}^{5}}{32\pi^{3}}
\int_{0}^{y_{max}(x)} \left\lbrace \frac{\alpha(y)}{M_{P}^{2}}y+ \right. \nonumber \\
& & 2(1-2x)[y_{max}(x)-y]\beta_{++}(y)+  \nonumber \\
& & \left. \gamma(y)y[2x_{max}-4x+y] \right\rbrace , \label{eq:vecelec}
\end{eqnarray}
where the following definitions were made
\begin{eqnarray}
\alpha(q^{2}) & = & \vert f(q^{2}) \vert^{2} + \lambda \vert g(q^{2}) \vert^{2} , \nonumber \\
\gamma(q^{2}) & = & 2f(q^{2})g(q^{2}) ,\nonumber \\
\beta_{++}(q^{2}) & = & \frac{1}{4M_{V}} \left\lbrace
\vert f(q^{2}) \vert^{2} + \lambda \vert a_{+}(q^{2}) \vert^{2}
-4M_{V}^{2}q^{2}\vert g(q^{2}) \vert^{2} \right. \nonumber \\
& & \left. +2(M_{P}^{2}-M_{V}^{2}-q^{2}) f(q^{2}) a_{+}(q^{2}) \right\rbrace , \nonumber \\
\lambda(q^{2}) & = & (M_{P}^{2}-M_{V}^{2} - q^{2})^{2} -4M_{V}^{2}q^{2} , \nonumber \\
x_{max} & = & \frac{M_{P}^{2}-M_{V}^{2}}{2M_{P}^{2}} \nonumber
\end{eqnarray}
Note that all of these expressions assume that lepton mass $m_{\ell}$ is zero.

Our model can be used to calculate all of these form factors.  In all the 
following expressions the inital meson is composed of quarks with masses $m_{1}$ and
$m_{2}$ and the final meson is composed of quarks with masses $m_{3}$ and $m_{2}$ 
(\emph{i.{}e.}~$m_{1}$ is the mass of the quark which decays to a new quark with
mass $m_{3}$, and $m_{2}$ is the mass of the spectator).  

The form factors for decay to a pseudoscalar meson are
\begin{eqnarray} 
f_{+}(q^{2}) & = & g^{}_{P} g^{}_{P'} \frac{9 \Lambda_{P}^{4} 
\Lambda_{P'}^{4}}{\pi^{2}} \nonumber \\
& & \int\!\! D\vec{x} \, x_{1} x_{2} \frac{\chi_{+}-\frac{3}{4}\bar{\Delta}(\sigma_{1}
+\sigma_{2})+\bar{\Delta}}{\bar{\Delta}^{5}}, \label{eq:fpff}
\end{eqnarray}
\begin{eqnarray} 
f_{-}(q^{2}) & = & g^{}_{P} g^{}_{P'} \frac{9 \Lambda_{P}^{4} 
\Lambda_{P'}^{4}}{\pi^{2}} \nonumber \\
& & \int\!\! D\vec{x} \, x_{1} x_{2} \frac{\chi_{-}-\frac{3}{4} \bar{\Delta} (\sigma_{1}
-\sigma_{2})}{\bar{\Delta}^{5}} .
\end{eqnarray}
The following definitions were made to simplify the expressions
\begin{eqnarray}
\int\!\! D\vec{x} & = & \int_{0}^{1} \left( \prod_{i=1}^{5} dx_{i} \right) \delta\!\!\left(
\sum_{i=1}^{5} x_{i} -1 \right) ,\nonumber \\
\Delta & = & \Lambda_{P'}^{2}x_{1}+ \Lambda_{P}^{2}x_{2} + m_{3}^{2} x_{3}
+m_{1}^{2}x_{4}+ m_{2}x_{5} ,\nonumber \\
\mu_{ij} & = & \frac{m_{i}}{m_{i}+m_{j}}, \nonumber \\
\sigma_{1,\,(2)} & = & x_{4,\,(3)}+\mu_{12,(23)} x_{2,\,(1)} ,\nonumber \\
\eta_{1,\,(2)} & = & x_{4,\,(3)}+\mu^{2}_{12,(23)} x_{2,\,(1)} ,\nonumber \\
\bar{\Delta} & = & \Delta + (\sigma_{1}^{2}-\eta_{1}+\sigma_{1}\sigma_{2}) M_{P}^{2}
\nonumber \\ 
& & +(\sigma_{2}-\eta_{2}+\sigma_{1}\sigma_{2}) M_{P'}^{2}-\sigma_{1}\sigma_{2}q^{2} , 
\nonumber \\
\kappa & = & m_{3}(m_{2}-m_{1})+m_{1}m_{2}+\frac{1}{2}(M_{P}^{2}+M_{P'}^{2}-q^{2}), 
\nonumber \\
\epsilon & = & (\sigma_{1}+\sigma_{2})\sigma_{1} M_{P}^{2}+
(\sigma_{1}+\sigma_{2})\sigma_{2} M_{P'}^{2}-\sigma_{1}\sigma_{2}q^{2} , \nonumber \\
\zeta_{1} & = & m_{1}m_{2}-\left[ \sigma_{1}\left( \sigma_{1}-1\right)+\sigma_{2}
\left( \sigma_{1}-\frac{1}{2} \right) \right]
M_{P}^{2} \nonumber \\
& & -\sigma_{2} \left( \sigma_{1}+\sigma_{2}-\frac{1}{2} \right) M_{P'}^{2}
 +\sigma_{2}\left( \sigma_{1}-\frac{1}{2} \right) q^{2} , \nonumber \\
\zeta_{2} & = & m_{2}m_{3}-\sigma_{1} \left( \sigma_{1}+\sigma_{2}-\frac{1}{2} \right) 
M_{P}^{2}
\nonumber \\
& & -\left[ \sigma_{2} \left( \sigma_{2}-1 \right)+\sigma_{1} \left( \sigma_{2}-\frac{1}{2}
\right) \right] M_{P'}^{2} \nonumber \\
& & +\sigma_{1} \left( \sigma_{2}-\frac{1}{2} \right) q^{2} , 
\nonumber \\
\chi_{\pm} & = & (\epsilon-\kappa)(\sigma_{1}\pm \sigma_{2})\pm \zeta_{1} 
+ \zeta_{2}  . \nonumber 
\end{eqnarray}
These definitions (in addition to $\alpha$ and $\mu$) will be used throughout the rest of 
this paper, with the obvious substitution of $M_{V}$ and $\Lambda_{V}$ for $M_{P'}$ and 
$\Lambda_{P'}$ when the final state is a vector meson. 

The form factors for decays to vector mesons are given by
\begin{eqnarray} 
g(q^{2}) & = & g^{}_{M_{P}} g^{}_{M_{V}} \frac{9 \Lambda_{P}^{4} \Lambda_{V}^{4}}{\pi^{2}}
\int\!\! D\vec{x} x_{1} x_{2} \nonumber \\
& &  \frac{ \sigma_{2}(m_{2}-m_{3}) + \sigma_{1} (m_{2}-m_{1})-m_{2} }{\bar{\Delta}^{5}}, 
\label{eq:gff}
\end{eqnarray}
\begin{eqnarray} 
f(q^{2}) & = & g^{}_{M_{P}} g^{}_{M_{V}} \frac{18 \Lambda_{P}^{4} \Lambda_{V}^{4}}{\pi^{2}}
\int\!\! D\vec{x} x_{1} x_{2} \times \nonumber \\
& & \frac{1}{\bar{\Delta}^{5}} \left[
m_{1}m_{2}m_{3}-\frac{1}{4}(m_{2}-m_{1}-2m_{3})
\bar{\Delta} \right. \nonumber \\
& & \left. + [\xi_{1}+\xi_{3}]M_{P}^{2} + [\xi_{2}+\xi_{3}]M_{V}^{2}
-\xi_{3} q^{2} \right] ,
\end{eqnarray}
\begin{eqnarray} 
a_{\pm}(q^{2})& = & g^{}_{M_{P}} g^{}_{M_{V}} \frac{18 \Lambda_{P}^{4} 
\Lambda_{V}^{4}}{\pi^{2}}
\int\!\! D\vec{x} x_{1} x_{2} \nonumber \\
& & \frac{\frac{1}{2} (\beta_{1} \pm \beta_{2} \mp \sigma_{2}
m_{3})}{\bar{\Delta}^{5}} . \label{eq:aff}
\end{eqnarray}
The following further definitions have been made,
\begin{eqnarray}
\xi_{1} & = & \sigma_{1}(m_{3}-m_{2})(1-\sigma_{1})-m_{1}\sigma_{1}^{2}, \nonumber \\
\xi_{2} & = & \sigma_{2}(m_{1}-m_{2})(1-\sigma_{2})-m_{3}\sigma_{2}^{2}, \nonumber \\
\xi_{3} & = & m_{2} \left[ \frac{1}{2}-\frac{1}{2}(\sigma_{1}+\sigma_{2})+
\sigma_{1}\sigma_{2} \right] +m_{1}\sigma_{1} \left( \frac{1}{2}-\sigma_{2} \right)
\nonumber \\
& & +m_{3}\sigma_{2} \left( \frac{1}{2} - \sigma_{1} \right) ,\nonumber \\
\beta_{1} & = & 2\sigma_{1} [m_{1}\sigma_{1}+m_{2}(1-\sigma_{1})], \nonumber \\
\beta_{2} & = & m_{2}(\sigma_{1}+\sigma_{2}-1-2\sigma_{1}\sigma_{2})-
m_{1}\sigma_{1}(1-2\sigma_{2}). \nonumber
\end{eqnarray}
%To calculate the form factors over the entire range of $q^{2}$ the following procedure
%was used. The form factor of interest was evaluated at eleven values of $q^{2}$ 
%from $q^{2}=0$ to $q^{2}=q_{max}^{2}=(M_{P}-M_{P',V})^{2}$  
%and these values were then fit by a cubic polynomial.
%This gave simple expressions for the various form factors, which were
%used in (\ref{eq:dgdyps}), (\ref{eq:pselec}), (\ref{eq:dgdyvec}) and (\ref{eq:vecelec}) to 
%get lepton spectra and total rates.
%
Excluding the $B_{c}$ decays
a total of sixteen pseudoscalar to pseudoscalar decays were considered.
Due to the difficulty with confinement the corresponding number of pseudoscalar to
vector decays that could be treated was only four.  
Table~\ref{drpred} shows the predictions for the decay rates and branching ratios
for all of the decays considered.  The values of the CKM matrix elements, and the 
necessary lifetimes were taken from~\cite{PDB}.

Many of the decay rates treated in this section have been measured, hence
most of the predictions can be compared to observed quantities.  Table~\ref{drpred}
shows the predicted and measured results for the branching ratios.  The 
experimental results are taken from~\cite{PDB} and the errors in the predictions 
represent the uncertainties in the CKM matrix elements.
Overall, the agreement with experiment is reasonable which increases
the level of confidence in the areas where direct comparison with experiment is not
possible.

Table~\ref{psthcomp} shows values of $f_{+}(0)$ as computed in this work and 
in various other
theoretical approaches.  The other approaches are widely 
varied:~\cite{demchukISGW} uses the ISGW model,~\cite{WBS} uses the WBS 
model,~\cite{sadzikowski} gives results from a bag model,~\cite{ivanovetalsdeq} uses a 
Dyson-Schwinger equation approach 
and~\cite{latticerev} gives lattice QCD results.  Of particular
interest is the work of Ivanov \emph{et al.}~\cite{heavyQCM3} which uses the quark 
confinement model.  This quark--meson model is based on similar 
considerations to the model used in this work
so its predictions should be close to ours.  

As well, the
decay $B \to D + \ell^{+} + \nu_{\ell}$ can be treated in a model independent 
way using the 
HQET.  Ivanov \emph{et~al.}~ have shown in several papers~\cite{QCMbook,
ivanovsant,heavyQCM2} that quark--meson models of the
type used here give the correct tree level HQET relations in the
infinite mass limit.  Nevertheless a direct check with finite quark mass is
useful.  The HQET gives the prediction~\cite{HQETrev}
$$
f_{+}^{HQET}(q^{2}_{max})=\frac{m_{B^{0}}+m_{D^{-}}}{2\sqrt{m_{B^{0}}m_{D^{-}}}}=1.138,
$$
which compares well with our value $f_{+}(q^{2}_{max})=1.133 .$

The most important comparison that can be made is with Ref.\cite{ivanovsant}.  This paper 
uses a
different vertex function to treat B and D decays.  This serves as a check on the 
dependence of the model on the choice of vertex function.  Apart from the case
$B \to \pi + \ell + \nu$ agreement with~\cite{ivanovsant} is very good.  
In addition~\cite{ivanovsant}
presents the values of $f_{+}(q^{2})$ over the full range of $q^{2}$.
Overall agreement is
good between the two calculations, Fig.~\ref{comp1} illustrates the agreement in 
the case
$D^{0} \to K^{-} + \ell^{+} + \nu_{\ell}$.  Fig.~\ref{comp2} shows the case 
$B^{0} \to \pi^{-} + \ell^{+} + \nu_{\ell}$, for which the agreement is better over
the whole range than indicated in Table~\ref{psthcomp}.

Due to lack of confinement very few pseudoscalar to vector decays can be calculated.
Of the few decays treated in this work only the decay
$B \to D^{*} + \ell + \nu_{\ell}$ has been studied extensively. Table~\ref{vcomp} 
compares our predictions with some other calculations. Overall the agreement is
reasonable.

\section{Semileptonic Decays of the $B_{c}$ Meson} \label{sec4}
The methods of the previous section can be directly applied to the semileptonic $B_{c}$
decays.  Using the procdeure outline above, decay rates, lepton spectra, and branching 
ratios 
can be computed.  In this work the lifetime of the $B_{c}$ was taken to be $0.5$ ps, 
which
agrees with the CDF value of $\tau_{B_{c}}^{CDF}=0.46^{+0.18}_{-0.16}\pm 0.03$ ps~
\cite{CDF}.
Table~\ref{tbc1} shows $f_{+}(0)$, $f_{+}(q_{max}^{2})$, the total decay rate
$\Gamma$ and the branching ratio for the four pseudoscalar decays.  
For the decays to vector mesons, values of the form factors at
$q^{2}=0$ as well as total decay rates and branching ratios
are displayed in Tables~\ref{tbc3} and~\ref{tbc4}.
%The full form factors for the decays
%to pseudoscalar mesons are displayed
%in Figs.~\ref{psff1} and~\ref{psff2}.  The form factors for the decays to vector
%mesons are displayed in Figs.~\ref{nvff1} to~\ref{pvff2}.

There are a number of other calculations of the semileptonic decays of $B_{c}$.
A comparison of some results for the dominant decay modes
is given in Table~\ref{bccomp}. In contrast to the situation
in Sect.~\ref{sec3} where our quark-meson model predictions, for the most part, 
agreed with other models and the various other models agreed with each other,
there are substantial differences between calculations of $B_{c}$ decays.
The clearest examples of this are the predictions for the decays to the $B_{s}^{*}$ and
$J/\psi$.  These two decays are expected to be the most important
semileptonic decay channels 
However there is disagreement not only over the values of the branching
ratios but also as to which decay will be favoured. For example, the 
quark-meson model used in this work predicts
the decay to the $J/\psi$ to be slightly favoured  over the 
decay to the $B_{s}^{*}$
while the heavy quark approach used in~\cite{sanchislozano} and ~\cite{colangelodefazio2}
predicts the decay to $B_{s}^{*}$ to dominate.
This divergence of predictions may be expected; the heavy--heavy quark content
of the $B_{c}$ poses a challenge for models. Light--quark mesons may
be constrained by chiral symmetry and heavy--light mesons by heavy quark symmetry.
On the other hand the physics of heavy--heavy systems is less constrained by
symmetries so extending models into this domain provides a severe test.

\section{Electromagnetic Decays $V \to P + \gamma$} \label{sec5}

In addition to semileptonic decays the electromagnetic decays of vector mesons can be
treated within our effective quark-meson coupling model.
Since the amplitude involves the matrix element
$\langle V | V^{\mu} | P \rangle$ it is clear this process will be related to
the form factor $g(q^{2})$. The
the amplitude for this process is
\begin{equation}
A=-2e\epsilon^{\mu \nu \alpha \beta} \tilde{\epsilon}_{\mu} \epsilon_{\nu}
p^{}_{V\, \alpha} p^{}_{P\, \beta} [Q_{1}g_{1}(0) + Q_{2} g_{2}(0)], 
\end{equation}
where $Q_{1,\,(2)}$ is the charge of $q_{1,\,(2)}$, and $p^{}_{P,\,V}$ are the momenta
of the pseudoscalar and vector mesons.  The functions $g_{i}(0)$ are the form factors
given by (\ref{eq:gff}), with the appropriate masses inserted, and with $q^{2}=0$.  
The appropriate masses in these functions are given by the interchange of $M_{P}$ and 
$M_{V}$ and the subscript which denotes which of the quark lines the gauge field is 
coupled to (\emph{i.{}e.}~ for $g_{1}(0)$ the appropriate expression sets 
$m_{3}$=$m_{1}$).
Defining
$g^{}_{VP \gamma}=2[Q_{1}g_{1}+Q_{2}g_{2}]$, and summing over initial and final 
polarizations gives
\begin{equation}
\vert A \vert^{2} =\frac{2\alpha \pi}{3} M_{V}^{4} \left[ 1-\frac{M_{P}^{2}}{M_{V}^{2}}
\right]^{2} g_{VP\gamma}^{2}, 
\end{equation}
where $\alpha=\frac{1}{137}$ is the fine structure constant.  Standard 
techniques\cite{peskinbook} yield
the total rate
\begin{equation} \label{eq:totrat}
\Gamma_{VP\gamma}=\frac{\alpha}{24} M_{V}^{3} g_{VP\gamma}^{2} 
\left( 1-\frac{M_{P}^{2}}{M_{V}^{2}} \right)^{3}.
\end{equation}

Electromagetic decays have been the subject of several theoretical studies.  
As well the  decay $J/\psi \to \eta_{c} + \gamma$ has been measured.  
Table~\ref{gvpg1} shows our predictions for $g^{}_{VP\gamma}$ along with the single 
experimental result and the predictions of some other models.  In~\cite{doschnarison} 
and~\cite{colangelodefazionardulli} two different heavy quark approaches were used.
The quark confinement model~\cite{heavyQCM1}, 
which has some similarity to the quark-meson model used in this work, 
gives results which are quite close to ours.

There are measured branching ratios for the $D^{*}$ decays, however no lifetime
measurement has been made.  Therefore our predictions (which do not include the 
lifetime) cannot be compared directly with experiment.  In order to obtain
branching ratios a theoretical estimate of the lifetime must be used.  The 
quark confinement model is the
ideal choice, since its predictions are closest to our work.  Using the results
from~\cite{heavyQCM1} and our predictions for the total rates (obtained from
(\ref{eq:totrat})) the following branching ratios are obtained:
\begin{eqnarray}
BR\left[ (D^{*})^{0}\to D^{0} + \gamma \right] & = & 33.0\% ,\nonumber \\
BR\left[ (D^{*})^{+}\to D^{+} + \gamma \right] & = & 1.43\%  . \nonumber
\end{eqnarray}
These compare well with the experimental values~\cite{PDB}
\begin{eqnarray}
BR_{expt.}\left[ (D^{*})^{0}\to D^{0} + \gamma \right] & = & 38.1\%, \nonumber \\
BR_{expt.}\left[ (D^{*})^{+}\to D^{+} + \gamma \right] & = & 1.1\%  . \nonumber
\end{eqnarray}

In order to treat the electromagnetic decay $B_{c}^{*} \to B_{c} + \gamma$ 
the mass of the $B_{c}^{*}$ meson must be specified.
Theoretical estimates~\cite{gershteinetal} indicate that the mass
difference should be small; $M_{B_{c}^{*}}-M_{B_{c}} < 100$ MeV. To
examine the effect of a small change in the $B_{c}^{*}$ mass, 
the self energy and coupling constant were calculated over a range
of masses.  These results were used to calculate $g^{}_{B_{c}^{*} B_{c} 
\gamma}$ and are displayed in Table~\ref{vbc1}.  The decay rate is
shown in Fig.~\ref{vbc2}.

The radiative decay of $B_{c}^{*}$ has not been studied extensively.
A QCD sum rule approach~\cite{alievetal}, 
using $M_{B^{*}_{c}}=6.6$ GeV and $M_{B_{c}}=6.3$ GeV, gives the result
$g^{SR}_{B^{*}_{c}B_{c}\gamma}=0.270 \pm 0.095 GeV^{-1}$.  Using these masses
our prediction is $g^{}_{B^{*}_{c}B_{c}\gamma}=0.2196 GeV^{-1}$.  The two
values are in agreement.

\section{Conclusion} \label{sec6}

A Lagrangian which models mesons in terms of an effective non-local quark-meson 
interaction
vertex~\cite{ENJL1,ivanovsant} was extended in this paper to describe mesons 
such as $B_{c}^{*}$ composed of two heavy quarks.
The model has the advantage of treating all quarks (heavy and
light) on the same footing, thereby permitting a unified investigation of 
heavy $\to$ heavy, and heavy $\to$ light quark decays. 
The model does not provide a complete dynamical description of quark interactions, 
meson masses can not be calculated and must be introduced as input parameters.
Quark masses and the parameters associated with the quark-meson vertex were determined 
by fitting the pseudoscalar and vector meson decay constants $f_{P}$ and $f_{V}$.
Due to lack of confinement in this approach some light vector mesons had to be
excluded from the analysis.

To both test the model, and demonstrate its versatility, a large number of 
semileptonic decays of $K$, $D$, and $B$ mesons were analysed.  Agreement with 
measured results and other theoretical approaches was good.  

The main focus of this work was on the analysis of the semileptonic decays of the 
doubly-heavy $B_{c}$ meson.  The form factors characterizing the strong interactions
of the $B_{c}$ system were computed over the entire available range of momentum
transfer.  Using these results, decay rates and branching ratios were computed for all 
eight decay going to mesonic ground states.

A comparison with some other approaches highlighted the significant differences among
various model predictions concerning the $B_{c}$.  On very general grounds one would 
expect the decays to $J/\psi$ and $(B^{*}_{s})^{0}$ states to be the most important. 
However, there is no agreement from models which of the channels dominates and
absolute rates differ by a factor of 3 to 4. This is in contrast to the situation 
in K, D and B meson decays where much smaller differences between different models
are found. 

As a further illustration of the versatility of the model, the electromagnetic
decays $V \to P + \gamma$ were investigated. A reasonable description
of $J/\psi$ and $D^*$ radiative decay was found and the rate for  
$B_{c}^{*} \to B_{c} + \gamma$
was calculated for a range of $B_{c}^{*}$ masses.

There is room for further analysis within the model considered in this work.
Hadronic decays, such as 
$B_{c}^{+} \to J/\psi + \pi^{+}$ can also be treated.
A detailed analysis of all of these decays, combined with the results for the leptonic
and semileptonic decays, could be used to make a prediction for the lifetime 
$\tau^{}_{B_{c}}$.  A further area that needs work is the difficulty with confinement.
The natural solution to this problem appears to be provided by the quark confinement
model~\cite{QCMbook}. Recently
Ivanov \emph{et~al.} have proposed a modification of the quark confinement model
which may aid its application to heavy mesons~\cite{newQCM}.

\vspace{1cm}
\noindent
\emph{Acknowledgements.}  We would like to thank M.A. Ivanov and P. Santorelli for a helpful
communication.  This work was supported in part by the Natural Sciences
and Engineering Research Council of Canada.

\newpage

\begin{table}[!ht]
\caption{Pseudoscalar Decay Constants}
\label{fittedps}
\begin{center}
\begin{tabular}{|c|c|c|}
\hline
Meson (P) & $f_{P}$ (MeV) & Source \\
\hline\hline
$\pi^{+}$ & $130.7\pm 0.5$ & Expt.~\cite{PDB}\\
$K^{+}$ & $159.8 \pm 1.8$ & Expt.~\cite{PDB}\\
$D^{+}$ & $183^{+12+41+9}_{-13-0-25}$ & Lattice QCD~\cite{latticefP} \\
$D^{+}_{s}$ & $229^{+10+51+3}_{-11-0-19}$ & Lattice QCD~\cite{latticefP} \\
$B^{+}$ & $156^{+12+29+9}_{-14-0-9}$ & Lattice QCD~\cite{latticefP}\\
$B^{0}_{s}$ & $177^{+11+39+13}_{-12-0-11}$ & Lattice QCD~\cite{latticefP}\\
\hline
\end{tabular}
\end{center}
\end{table}

\begin{table}[!ht]
\caption{Fitted Values of Pseudoscalar Meson Properties} 
\label{pcalc}
\begin{center}
\begin{tabular}{|c|c|c|c|}
\hline
Meson  & $\Pi(M^{2})$ ($\textrm{GeV}^{2}$) & $g^{}_{P}$ & $f_{P}$ (MeV) \\
\hline\hline
$\pi^{+}$ & 0.038588 & 5.14916 & 131.06 \\
$K^{+}$ & 0.068760 & 5.16139 & 160.85 \\
$D^{+}$ & 0.083122 & 6.26671 & 182.80 \\
$D_{s}^{+}$ & 0.087025 & 6.82039 & 223.81 \\
$B^{+}$ & 0.134842 & 6.44510 & 142.21 \\
$B_{s}^{0}$ & 0.136072 & 7.75408 & 187.42 \\
\hline
\end{tabular}
\end{center}
\end{table}

\begin{table}[!ht]
\caption{Fitted Values of Vector Meson Properties} 
\label{vcalc}
\begin{center}
\begin{tabular}{|c|c|c|c|}
\hline
Meson & $\Pi(M^{2})$ ($\textrm{GeV}^{2}$) & $g^{}_{V}$ & $f_{V}$ \\
\hline\hline
$J/\psi$ & 0.095419 & 8.81505 & 0.131227 \\
$\Upsilon$ & 0.621464 & 7.54393 & 0.074796 \\
\hline
\end{tabular}
\end{center}
\end{table}

\begin{table}[!ht]
\caption{Predictions for Decay Rates and Branching Ratios} 
\label{drpred}
\begin{center}
\begin{tabular}{|c|c|c|c|c|}
\hline
Parent & Daughter & $\Gamma$ ($ps^{-1}$) & $BR(\%)$ & $BR_{expt.}(\%)$ \\
\hline\hline
$K^{+}$ & $\pi^{0}$ & $3.589 \times 10^{-6}$ & $4.45$ & $3.18 \pm 0.08 $ \\
$K_{S}^{0}$ & $\pi^{-}$ & $7.274 \times 10^{-6}$ & $0.0650$ & $0.0670\pm 0.0007 $ \\
$K_{L}^{0}$ & $\pi^{-}$ & $7.274 \times 10^{-6}$ & $37.6$ & $38.78 \pm 0.27$\\
$D^{0}$ & $\pi^{-}$ & $5.488 \times 10^{-3}$ & $0.228$ & $0.37 \pm 0.06 $ \\
$D^{0}$ & $K^{-}$ & $8.476 \times 10^{-2}$ & $3.52$ & $3.50 \pm 0.17 $\\
$D^{+}$ & $\pi^{0}$ & $2.790 \times 10^{-3}$ & $0.295$ & $0.31 \pm 0.15 $\\
$D^{+}$ & $K^{0}$ & $8.515 \times 10^{-2}$ & $9.000$ & $6.8 \pm 0.8 $\\
$D_{s}^{+}$ & $K^{0}$ & $4.184 \times 10^{-3}$ & $0.195$ &\\
$D_{s}^{+}$ & $D^{0}$ & $4.786 \times 10^{-8}$ & $2.24 \times 10^{-6}$& \\
$B^{0}$ & $\pi^{-}$ & $6.455 \times 10^{-5}$ & $0.0101$ & $0.018 \pm 0.006 $\\
$B^{0}$ & $D^{-}$ & $1.716 \times 10^{-2}$ & $2.68$ & $2.00 \pm 0.006 $\\
$B^{0}$ & $(D^{*})^{-}$ & $3.983 \times 10^{-2}$ & $6.21$ & $4.60 \pm 0.27 $\\
$B^{+}$ & $\pi^{0}$ & $3.457 \times 10^{-5}$ & $5.70 \times 10^{-3}$ & $<0.22 $\\
$B^{+}$ & $D^{0}$ & $1.726 \times 10^{-2}$ & $2.85$ & $1.86 \pm 0.33 $ \\
$B^{+}$ & $(D^{*})^{0}$ & $4.059 \times 10^{-2}$ & $6.70$ & $5.3 \pm 0.8 $ \\
$B_{s}^{0}$ & $K^{-}$ & $6.118 \times 10^{-5}$ & $9.42 \times 10^{-3}$ &\\
$B_{s}^{0}$ & $D_{s}^{-}$ & $1.642 \times 10^{-2}$ & $2.53$ &\\
$B_{s}^{0}$ & $(D_{s}^{*})^{-}$ & $4.185 \times 10^{-2}$ & $6.45$ &\\
$B_{s}^{0}$ & $B^{-}$ & $2.619 \times 10^{-8}$ & $4.03 \times 10^{-6}$ &\\
$B_{s}^{0}$ & $(B^{*})^{-}$ & $8.032 \times 10^{-10}$ & $1.24 \times 10^{-7}$& \\
\hline
\end{tabular}
\end{center}
\end{table}

%\begin{table}[!ht]
%\caption{Predicted and Experimental Branching Ratios}
%\label{expcomp}
%\begin{center}
%\begin{tabular}{|c|c|c|c|}
%\hline
%Parent & Daughter & Predicted BR(\%) & Expt. BR(\%) \\
%\hline\hline
%$K^{+}$ & $\pi^{0}$ & $4.45 \pm 0.09$ & $3.18 \pm 0.08$ \\
%$K_{S}^{0}$ & $\pi^{-}$ & $0.0650 \pm 0.0014$ & $0.0670 \pm 0.0007$ \\
%$K_{L}^{0}$ & $\pi^{-}$ & $37.6 \pm 0.78$ & $38.78 \pm 0.27$ \\
%$D^{0}$ & $\pi^{-}$ & $0.228 \pm 0.033$ & $0.37 \pm 0.06$ \\
%$D^{0}$ & $K^{-}$ & $3.52 \pm 0.01$ & $3.50 \pm 0.17$ \\
%$D^{+}$ & $\pi^{0}$ & $0.295 \pm 0.042$ & $0.31 \pm 0.15$ \\
%$D^{+}$ & $K^{0}$ & $9.00 \pm 0.02$ & $6.8 \pm 0.8$ \\
%$B^{0}$ & $\pi^{-}$ & $0.0101 \pm 0.0030$ & $0.018 \pm 0.006$ \\
%$B^{0}$ & $D^{-}$ & $2.68 \pm 0.22$ & $2.00 \pm 0.25$ \\
%$B^{0}$ & $(D^{*})^{-}$ & $6.21 \pm 0.52$ & $4.60 \pm 0.27$ \\
%$B^{+}$ & $\pi^{0}$ & $(5.70 \pm 1.72) \times 10^{-3}$ & $<0.22$ \\
%$B^{+}$ & $D^{0}$ & $2.85 \pm 0.24$ & $1.86 \pm 0.33$ \\
%$B^{+}$ & $(D^{*})^{0}$ & $6.70 \pm 0.56$ & $5.3 \pm 0.8$ \\
%\hline
%\end{tabular}
%\end{center}
%\end{table}

\begin{table}[!ht]
\caption{Comparison of 
this work with other
approaches for the form factor $f_{+}(0)$. 
Here P is the parent meson and D is the Daughter} 
\label{psthcomp}
\begin{center}
\begin{tabular}{|c|c|c|c|c|c|c|c|c|c|c|}
\hline
P & D & This work & \cite{WBS} & \cite{demchukISGW} & \cite{sadzikowski} & \cite{heavyQCM3} 
& \cite{ivanovsant} & \cite{latticerev} & \cite{ivanovetalsdeq}\\
\hline\hline
$K^{+}$ & $\pi^{0}$ & 0.9617  & & & & & 0.98 & & \\
$D^{0}$ & $K^{-}$ & 0.7869  & 0.76 & 0.780 & 0.71 & 0.78 & 0.74 & 0.73 & 0.79\\
$D^{0}$ & $\pi^{-}$ & 0.6292  & 0.69 & 0.681 & 0.8 & & & 0.65 & 0.86\\
$B^{0}$ & $D^{-}$ & 0.7977  & 0.69 & 0.684 & & & 0.73 & & 0.65\\
$B^{0}$ & $\pi^{-}$ & 0.2848  & 0.33 & 0.293 & 0.33 & 0.53 & 0.51 & 0.27 & 0.43\\
$B_{s}^{0}$ & $K^{-}$ & 0.2452  & & & 0.36 & & & & \\
\hline
\end{tabular}
\end{center}
\end{table}

\begin{table}[!ht]
\caption{Vector Form Factors for the Decay $B \to D^{*} + \ell^{+} + \nu_{\ell}$} 
\label{vcomp}
\begin{center}
\begin{tabular}{|c|c|c|c|}
\hline
Reference & $g(0)$ ($GeV^{-1}$) & $a_{+}(0)$ ($GeV^{-1}$) & $f(0)$ ($GeV$) \\
\hline\hline
This Work & -0.10391 & -0.09240 & 5.286 \\
\cite{WBS} & -0.09745 & -0.09471 & 4.736 \\
\cite{scoraisgur} & -0.16 & -0.15 & 6.863 \\
\cite{ivanovetalsdeq} & -0.09054 & -0.07271 & 3.863 \\
\hline
\end{tabular}
\end{center}
\end{table}

\begin{table}[!ht]
\caption{Predictions for $B_{c} \to P$ Decays}
\label{tbc1}
\begin{center}
\begin{tabular}{|c|c|c|c|c|}
\hline
P & $f_{+}(0)$ & $f_{+}(q_{max}^{2})$ & $\Gamma$ ($ps^{-1}$) & BR (\%) \\
\hline\hline
$B^{0}$ & 0.4504 & 0.6816 & $9.7001 \times 10^{-4}$ & 0.049 \\
$B_{s}^{0}$ & 0.5917 & 0.8075 & $1.8774 \times 10^{-2}$ & 0.94 \\
$D^{0}$ & 0.1446 & 1.017 & $2.8244 \times 10^{-5}$ & 0.0014 \\
$\eta_{c}$ & 0.5359 & 1.034 & $1.0355 \times 10^{-2}$ & 0.52 \\
\hline
\end{tabular}
\end{center}
\end{table}

\begin{table}[!ht]
\caption{Predictions for $B_{c} \to V$ Decays} 
\label{tbc3}
\begin{center}
\begin{tabular}{|c|c|c|}
\hline
V & $\Gamma$ ($ps^{-1}$) & BR(\%) \\
\hline\hline
$(B^{*})^{0}$ & $1.048 \times 10^{-3}$ & 0.052 \\
$(B_{s}^{*})^{0}$ & $2.872 \times 10^{-2}$ & 1.44 \\
$(D^{*})^{0}$ & $4.739 \times 10^{-5}$ & 0.0024 \\
$J/\psi$ & $2.943 \times 10^{-2}$ & 1.47 \\
\hline
\end{tabular}
\end{center}
\end{table}

\begin{table}[!ht]
\caption{Predictions for the form factors at $q^{2}=0$ for $B_{c} \to V$ decays.}
\label{tbc4}
\begin{center}
\begin{tabular}{|c|c|c|c|}
\hline
V & $g(0)$ $GeV^{-1}$ & $a_{+}(0)$ $GeV^{-1}$ & $f(0)$ $GeV$ \\
\hline\hline
$(B^{*})^{0}$ & -0.1671 & -0.0463 & 3.383 \\
$(B_{s}^{*})^{0}$ & -0.2402 & -0.0673 & 5.506 \\
$(D^{*})^{0}$ & -0.0211 & -0.0127 & 0.8296 \\
$J/\psi$ & -0.0784 & -0.0543 & 4.918 \\
\hline
\end{tabular}
\end{center}
\end{table}

\begin{table}[!ht]
\caption{Branching Ratios for the Semileptonic Decays of the $B_{c}$.}
\label{bccomp}
\begin{center}
\begin{tabular}{|c|c|c|c|c|c|}
\hline
Decay Meson & This Work & \cite{bcsr1} & \cite{sanchislozano} & \cite{bcsr2} &
\cite{colangelodefazio2} \\
\hline\hline
$B_{s}^{0}$ & $0.94$\% & $(0.68\pm0.23)$\% & $1.35$\% & & $0.80$\% \\
$(B_{s}^{*})^{0}$ & $1.44$\% & $(0.68\pm0.23)$\% & $3.22$\% & & $2.3$\% \\
$\eta_{c}$ & $0.52$\% & $(0.57\pm0.17)$\% & $0.553$\% & $0.836$\% & $0.15$\% \\
$J/\psi$ & $1.47$\% & $(0.68\pm0.17)$\% & $2.32$\% & $2.13$\% & $1.5$\% \\
\hline
\end{tabular}
\end{center}
\end{table}

\begin{table}[!ht]
\caption{The decay constant $g_{VP\gamma}$ in $GeV^{-1}$.} 
\label{gvpg1}
\begin{center}
\begin{tabular}{|c|c|c|c|c|c|c|}
\hline
V & This Work & Expt.~\cite{PDB} & \cite{ivanovetalsdeq} & \cite{doschnarison} & \cite{colangelodefazionardulli} & \cite{heavyQCM1} \\
\hline \hline
$(D^{*})^{0}$ & 2.0321 & & 1.043 & 1.079 & 1.598 & 1.939 \\

$(D^{*})^{+}$ & 0.5224 & & 0.1535 & 0.0702 & 0.2886 & 0.3950 \\

$(D_{s}^{*})^{+}$ & 0.2369 & & & & 0.1917 & 0.2598 \\

$J/\psi$ & 0.7419 & 0.5538 & & & &  \\

$(B^{*})^{0}$ & 0.9770 & & 0.3098 & 0.4177 & 0.5720 & 0.9104 \\

$(B^{*})^{+}$ & 1.7627 & & 0.3461 & 0.6540 & 0.9701 & 1.618 \\

$(B_{s}^{*})^{0}$ & 0.6417 & & & & &  \\

$\Upsilon$ & 0.1314 & & & & &  \\
\hline
\end{tabular}
\end{center}
\end{table}

\begin{table}[!ht]
\caption{Calculations of $B_{c}^{*}$ Properties and $g_{B_{c}^{*} B_{c}
\gamma}$.} 
\label{vbc1}
\begin{center}
\begin{tabular}{|c|c|c|c|}
\hline
$M_{B_{c}^{*}}$ ($GeV$) & $\Pi \left( (M_{B_{c}^{*}})^{2} \right)$ ($GeV^{2}$) & 
$g_{B_{c}^{*}}$ & $g_{B_{c}^{*} B_{c} \gamma}$ ($GeV^{-1}$)\\
\hline \hline
6.25 & 0.2843 &	8.627 & 0.3013 \\

6.26 & 0.2877 &	8.541 & 0.3011 \\

6.27 & 0.2912 & 8.455 & 0.3008 \\

6.28 & 0.2947 &	8.369 & 0.3006 \\

6.29 & 0.2984 &	8.283 & 0.3003 \\

6.30 & 0.3021 & 8.196 & 0.3001 \\

6.31 & 0.3059 & 8.109 & 0.2999 \\

6.32 & 0.3097 & 8.022 & 0.2995 \\

6.33 & 0.3137 & 7.934 & 0.2993 \\

6.34 & 0.3178 & 7.846 & 0.2990 \\

6.35 & 0.3220 & 7.758 & 0.2987 \\
\hline
\end{tabular}
\end{center}
\end{table}

\clearpage

\begin{figure}[!ht]
\begin{center}
\input{comp1.tex}
\end{center}
\caption{Form factor for the decay $D^{0} \to K^{-}$
calculated in this work (dashed line)
and in Ref.~\cite{ivanovsant} (solid line).}
\label{comp1}
\end{figure}

\begin{figure}[!ht]
\begin{center}
\input{comp2.tex}
\end{center}
\caption{Form factor for the decay $B^{0} \to \pi^{-}$
calculated in this work (dashed line)
and in Ref.~\cite{ivanovsant} (solid line).}
\label{comp2}
\end{figure}

\begin{figure}[!ht]
\begin{center}
\input{emdec.tex}
\end{center}
\caption{The decay rate $\Gamma \left[ B_{c}^{*} \to B_{c} + \gamma \right]$
vs. $M_{B_{c}^{*}}$.} \label{vbc2}
\end{figure}

\end{document}